\documentstyle[aps,preprint,epsfig]{revtex}
\def\bge{\begin{equation}}
\def\ene{\end{equation}}
\def\bg{\begin{eqnarray}}
\def\en{\end{eqnarray}}

\def\ubar{{\bar{u}}}
\def\dbar{{\bar{d}}}

\def\cbar{{\bar{c}}}
\def\d0bar{{\bar{D}^0}}

\def\vr{\vec{r}}
\def\vx{\vec{x}}

\def\e{\epsilon}

\begin{document}
%
\preprint{
\vbox{
\hbox{ADP-98-48/T317, OSUNT98-13}
}}

%
%

\title{Charmed Mesic Nuclei:\\ 
Bound $D$ and $\bar{D}$ states with $^{208}$Pb}
\author{K. Tsushima$^1$, D.H. Lu$^1$,
A.W. Thomas$^1$, K. Saito$^2$ and R.H. Landau$^3$}
\address {$^1$ Department of Physics and Mathematical Physics,
        and Special Research Center for the
        Subatomic Structure of Matter,
        University of Adelaide, Adelaide, 5005 Australia}
\address{$^2$ Physics Division, Tohoku College of Pharmacy,
                  Sendai 981-8558, Japan}
\address{$^3$ Department of Physics, Oregon State University,
        Corvallis, OR 97331, USA}

\date{\today}
\maketitle
\parshape=1 0.75in 5.5in \indent
{\small {\\
We show that the $D^-$ meson will form narrow bound states with
$^{208}$Pb. Mean field potentials for the $D^0$, $\d0bar$ and $D^-$
in $^{208}$Pb are calculated self-consistently using the quark-meson 
coupling (QMC) model in local density approximation. The 
meson-$^{208}$Pb bound state energies are then calculated by solving 
the Klein-Gordon equation with these potentials.
The experimental confirmation and comparison with the $\d0bar$ and $D^0$
will provide distinctive information
on the nature of the interaction between the charmed meson and matter.

\bigskip
PACS numbers: 21.30.Fe, 24.10.Jv, 14.40.Lb, 21.10.Dr, 11.30.Rd
\\}



%

In relativistic models of nuclear structure like QHD~\cite{qhd} or
QMC~\cite{gui,finite,finite1,saito,kaon,etao,blu,mue,jin}, the
isoscalar-scalar meson, $\sigma$, is responsible for a large reduction
in the mass of the nucleon ($m_N \to m_N^*$). Although there is no
firm evidence of a $\sigma$ meson with mass in
the range 500
-- 600 MeV (c.f. the Particle Data Group discussion of the 
$f_0(400 - 1200)$ and $\sigma$~\cite{pdata}),
there is considerable empirical justification for using the
$\sigma$ as a phenomenological representation of correlated
two-pion-exchange between nucleons~\cite{sigma}. In QMC the
justification for coupling the $\sigma$ to the confined light quarks
(hereafter referred to as $q$) is that dynamical chiral symmetry
breaking requires that the confined quarks couple to pions. Quarks in
different hadrons can interact by exchanging two pions and this is
represented phenomenologically by $\sigma$-exchange. Using this model
it is possible to investigate the reduction of the masses not only
of baryons, but also of mesons containing light
quarks~\cite{saito,kaon,etao}. 
(See Refs.~\cite{will,qhdomega,asa,qcd,cha,fri,kli} for other 
approaches not all of which result in a mass reduction for the mesons.) 

The result for the $\omega$-meson within relativistic mean field 
models is especially interesting. Because
it consists of almost pure $q$-$\bar{q}$ pairs (ideal mixing), in an
isoscalar nucleus the $q$ and $\bar{q}$ feel equal and opposite vector
potentials. This means that the effect of the mean $\sigma$-field is
unmasked which leads to expectations of quite deeply bound
$\omega$-nucleus states~\cite{etao,qhdomega,hayano,hayano2,weise}.
These states are currently the subject of intense experimental
investigation, with the most promising involving recoilless production
in the (d,$^3$He) reaction at GSI~\cite{hayano,hayano2}.

In this paper we consider a possibility that is in some ways even more
exciting, in that it promises more specific information on the
relativistic mean fields in nuclei and the nature of dynamical chiral
symmetry breaking. We focus on systems containing an anti-charm quark
and a light quark ($\bar{c}q$), which have no strong decay channels if
bound. If we assume that dynamical chiral symmetry breaking is the
same for the light quark in the charmed meson as in purely light-quark
systems, we expect the same $\sigma$-$q$ coupling constant. (On the
one hand, this assumption seems quite reasonable, given the success
in reproducing the pion coupling constants for various charmed baryons
using a model based on a common light quark pion coupling \cite{IK98},
while on the other hand, the present predictions could also be
viewed as an independent test of this picture.)
In the absence of any strong interaction, the $D^-$ will form
atomic states, bound by the Coulomb potential. We use the QMC model to
estimate the effect of the strong interaction. The resulting binding
for, say, the 1s level in $^{208}$Pb is between ten and thirty MeV and
should provide a very clear experimental signature.


Systems of the form $\bar{q}c$ are also extremely interesting because
the coupling of the mean vector field to the light anti-quark will be
attractive. Indeed, we expect a $D$ meson ($c\bar{q}$) to experience
an attraction in excess of 100 MeV in an average size atomic
nucleus. Unfortunately, the $D$ meson in matter will also couple
strongly to open channels such as $D N \rightarrow B_c (\pi's)$, with
$B_c$ a charmed baryon. Our present knowledge does not permit an
accurate calculation of the corresponding widths, which may be 10-100
MeV. At the lower end of this range, such states should be able to be
seen as very deeply bound mesic-nuclei, while at the upper end they
may not be detectable. We regard the widths as an experimental issue
at present and show the $^{208}_{\d0bar}$Pb bound state energies
without the effect of absorption.

A detailed description of the Lagrangian density and the
mean-field equations of motion needed to describe a finite nucleus
is given in Refs.~\cite{finite,finite1}. At position $\vr$ in a
nucleus (the coordinate origin is taken at the center of the nucleus),
the Dirac equations for the quarks and antiquarks in the $D$ and
$\bar{D}$ meson bags are given by~\cite{kaon,etao}:
\bg
\left[ i \gamma \cdot \partial_x - (m_q - V^q_\sigma(\vr))
\mp \gamma^0
\left( V^q_\omega(\vr) + \frac{1}{2} V^q_\rho(\vr) \right) \right]
\left(\begin{array}{c} \psi_u(x)\\ \psi_\ubar(x)\\ \end{array}\right)
 &=& 0,
\label{diracu}
\\
\left[ i \gamma \cdot \partial_x - (m_q - V^q_\sigma(\vr))
\mp \gamma^0
\left( V^q_\omega(\vr) - \frac{1}{2} V^q_\rho(\vr) \right) \right]
\left(\begin{array}{c} \psi_d(x)\\ \psi_\dbar(x)\\ \end{array} \right)
 &=& 0,
\label{diracd}
\\
\left[ i \gamma \cdot \partial_x - m_{c} \right]
\psi_{c} (x)\,\, ({\rm or}\,\, \psi_{\cbar}(x)) &=& 0.
\label{diracsc}
\en
The mean-field potentials for a bag centered at position $\vr$ in
the nucleus are defined by $V^q_\sigma(\vr) = g^q_\sigma
\sigma(\vr), V^q_\omega(\vr) = g^q_\omega \omega(\vr)$ and
$V^q_\rho(\vr) = g^q_\rho b(\vr)$, with $g^q_\sigma, g^q_\omega$ and
$g^q_\rho$ the corresponding quark and meson-field coupling
constants. (Note that we have neglected a possible, very slight
variation of the scalar and vector mean-fields inside the meson bag
due to its finite size~\cite{finite}.) The mean meson fields are
calculated self-consistently by solving Eqs.~(23) -- (30) of
Ref.~\cite{finite1}, namely, by solving a set of coupled 
non-linear differential equations for static, spherically symmetric nuclei,
resulting from the variation of the effective Lagrangian density 
involving the quark degrees of freedom and the scalar, vector and Coulomb  
fields in mean field approximation.

The normalized, static solution for the ground state quarks or antiquarks
in the meson bags may be written as:
\bg
\psi_f (x) = N_f e^{- i \epsilon_f t / R_j^*} \psi_f (\vx),
\qquad (j = D, \bar{D}),
\label{wavefunction}
\en
where $f = u,\ubar,d,\dbar,c,\cbar$ refers to quark flavors,
and $N_f$ and $\psi_f (\vx)$ are the normalization factor and
corresponding spin and spatial part of the wave function. The bag
radius in medium, $R_j^*$, which depends on the hadron species to 
which the quarks and antiquarks belong, will be determined through the
stability condition for the (in-medium) mass of the meson against the
variation of the bag radius~\cite{finite,finite1,kaon,etao}
(see also Eq.~(\ref{equil})). The eigenenergies $\e_f$ in
Eq.~(\ref{wavefunction}) in units of $1/R_j^*$ 
 are given by
\bg
\left( \begin{array}{c} \e_u(\vr) \\ \e_{\ubar}(\vr) \end{array} \right)
&=& \Omega_q^*(\vr) \pm R_j^* \left(
V^q_\omega(\vr) + \frac{1}{2} V^q_\rho(\vr) \right),
\label{uenergy}
\\
\left( \begin{array}{c} \e_d(\vr) \\ \e_{\dbar}(\vr) \end{array} \right)
&=& \Omega_q^*(\vr) \pm R_j^* \left(
V^q_\omega(\vr) - \frac{1}{2} V^q_\rho(\vr) \right),
\label{denergy}
\\
\e_{c}(\vr) &=& \e_{\cbar}(\vr) = \Omega_{c}(\vr),
\label{cenergy}
\en
where $\Omega_q^*(\vr) = \sqrt{x_q^2 + (R_j^* m_q^*)^2}$, with
$m_q^* = m_q - g^q_\sigma \sigma(\vr)$ and
$\Omega_{c}(\vr) = \sqrt{x_{c}^2 + (R_j^* m_{c})^2}$.
The bag eigenfrequencies, $x_q$ and $x_{c}$, are
determined by the usual, linear boundary condition~\cite{finite,finite1}.
(Note that the lowest eigenenergy value for the Dirac equation  
(Hamiltonian) for the quark, which is positive, should be identified with  
a constituent quark mass.)  

%
%
The $D$ and $\bar{D}$ meson masses
in the nucleus at position $\vr$, are calculated by:
\bg
%
m_j^*(\vr) &=& \frac{\Omega_q^*(\vr)
+ \Omega_c(\vr) - z_j}{R_j^*}
+ {4\over 3}\pi R_j^{* 3} B,
\label{md}
\\
& &\left. \frac{\partial m_j^*(\vr)}
{\partial R_j}\right|_{R_j = R_j^*} = 0,
\hspace{8em} (j = D, \bar{D}).
\label{equil}
\en
%
%
%
%
In Eq.~(\ref{md}), the $z_j$ parametrize the sum of the
center-of-mass and gluon fluctuation effects, and are assumed to be
independent of density. The parameters are determined in free space to
reproduce their physical masses.

In this study we chose the values $m_q \equiv m_u = m_d = 5$
MeV and $m_c = 1300$ MeV for the current quark masses, and $R_N = 0.8$
fm for the bag radius of the nucleon in free space. Other input
parameters and some of the quantities calculated are listed
in Table~\ref{bagparam}. The parameters at the hadronic level
associated with the core nucleus can be found in
Refs.~\cite{finite,finite1}. We stress that while the model has a
number of parameters, only three of them, $g^q_\sigma$, $g^q_\omega$
and $g^q_\rho$, are adjusted to fit nuclear data -- namely the
saturation energy and density of symmetric nuclear matter and the bulk
symmetry energy. None of the results for nuclear properties depend
strongly on the choice of the other parameters -- for example, the
relatively weak dependence of the final results for the properties of
finite nuclei, on the chosen values of the current quark mass and bag
radius, is shown explicitly in Refs.~\cite{finite,finite1}. Exactly
the same coupling constants, $g^q_\sigma$, $g^q_\omega$ and
$g^q_\rho$, are used for the light quarks in the mesons as in the
nucleon. However, in studies of the kaon system, we found that it was
phenomenologically necessary to increase the strength of the vector
coupling to the non-strange quarks in the $K^+$ (by a factor of
$1.4^2$) in order to reproduce the empirically extracted $K^+$-nucleus
interaction~\cite{kaon}. It is not yet clear whether this is a
specific property of the $K^+$, which is a pseudo-Goldstone boson, or
a general feature of the interaction of a light quark, perhaps
associated with the Pauli exclusion principle. In any case, we show
results for the $\bar{D}$ binding energies with both choices for this
potential, in order to test the theoretical uncertainty. The $\omega$
mean field potential with the larger coupling will be labelled
$\tilde{V}^q_\omega$ $(= 1.4^2 V^q_\omega)$.

Through Eqs.~(\ref{diracu}) -- (\ref{equil}) we self-consistently
calculate effective masses, $m^*_j(\vr)$ ($j=D,\bar{D}$),
and mean field potentials, $V^q_{\sigma,\omega,\rho}(\vr)$,
at position $\vr$ in the nucleus.
The scalar and vector  potentials felt by the hadron,
which will depend only on
the distance from the center of the nucleus, $r = |\vr|$, are given by:
\bg
V^j_s(r)
&=& m^*_j(r) - m_j,
\label{spot}\\
%
V^{D^-}_v(r) &=&
 V^q_\omega(r) - \frac{1}{2}V^q_\rho(r) - A(r),
\label{vdpot1}
\\
V^{\bar{D^0}}_v(r) &=&
 V^q_\omega(r) + \frac{1}{2}V^q_\rho(r),
\label{vdpot2}
\\
V^{D^0}_v(r) &=&
-\left( V^q_\omega(r) + \frac{1}{2}V^q_\rho(r) \right),
\label{vdpot3}
\en
where $A(r)$ is the Coulomb interaction between the meson and the nucleus.
Note that the $\rho$ meson mean field potential, $V^q_\rho(r)$, is negative
in a nucleus with a neutron excess, such as e.g., $^{208}$Pb.
For the larger $\omega$ meson coupling, suggested by $K^+A$ scattering,
$V^q_\omega(r)$ is replaced by $\tilde{V}^q_\omega(r)$.

Before showing the calculated potentials for the $D^-$ in $^{208}$Pb
which is particularly interesting in view of the strong Coulomb field,
we first show in Fig.~\ref{dmassmatter}
the mass shift of the $D (\bar{D})$ meson, calculated in symmetric
nuclear matter
(no contributions from the $\rho$ and Coulomb fields).
As is expected, the relation
$(m_{D,\bar{D}} - m^*_{D,\bar{D}}) \simeq
\frac{1}{3} (m_N - m^*_N)$ is well realized~\cite{saito}.

Next, in Fig.~\ref{dmespot} we show the sum of the potentials
for the $D^-$ in $^{208}$Pb for the two choices of
$V^{D^-}_s(r) + V^{D^-}_v(r)$ (the dashed line corresponds to
$\tilde{V}^q_\omega(r)$
and the dotted line to ${V}^q_\omega(r)$).
%
Because the $D^-$ meson is heavy and may be described well
in the (nonrelativistic) Schr\"{o}dinger equation, one
expects the existence of the $_{D^-}^{208}$Pb bound states
just from inspection of the naive sum of the potentials,
in a way which does not distinguish the Lorentz vector or
scalar character.

Now we are in a position to calculate the bound state energies for the
$D$ and $\bar{D}$ in nuclei,
using the potentials calculated in QMC.
There are several variants of the dynamical equation for a bound meson-nucleus
system. Consistent with the mean field picture of QMC, we actually
solve the Klein-Gordon equation:
%
\bg
[ \nabla^2 &+& (E_j - V^j_v(r))^2- m^{*2}_j(r) ]\,
\phi_j(\vr) = 0
\label{kgequation}
\en
where $E_j$ is the total energy of the meson
(the binding energy is $E_j-m_j$).
To deal with the long range Coulomb potential, we first expand the quadratic
term (the zeroth component of Lorentz vector) as ,
$(E_j - V^j_v(r))^2 = E_j^2 + A^2(r) + V^2_{\omega\rho}(r) +
2 A(r)V_{\omega\rho}(r) - 2 E_j (A(r)+V_{\omega\rho}(r))$,
where $V_{\omega\rho}(r)$ is the combined potential due to
$\omega$ and $\rho$ mesons
($V_{\omega\rho}(r) = V_\omega^q(r)-{1\over 2}V_\rho^q(r)$ for $D^-$).
Then Eq.~(\ref{kgequation}) can be rewritten as an effective
Schr\"odinger-like equation,
\begin{equation}
\left[ - {\nabla^2\over 2m_j} + V_j(E_j,r)\right] \Phi_j(r)
= {E_j^2-m_j^2 \over 2m_j}\Phi_j(r)
\label{schrodinger}
\end{equation}
where $\Phi_j(r) = 2m_j\phi_j(r)$ and
$V_j(E_j,r)$ is an effective energy-dependent
potential which can be split
into three pieces (Coulomb, vector and scalar parts),
\begin{equation}
V_j(E_j,r) = {E_j\over m_j} A(r) + {2E_jV^j_{\omega\rho}(r)
- (A(r)+V^j_{\omega\rho}(r))^2\over 2m_j} + {{m^*_j}^2(r)-m^2_j\over 2m_j}.
\end{equation}
Note that only the first term in this equation is  a long range
interaction and thus needs special treatment, the second and third
terms are  short range interactions.
In practice, Equation~(\ref{schrodinger}) is first converted into
momentum space representation via a Fourier transformation and is then
solved  using the Kwon-Tabakin-Land\'e technique\cite{landau}.
We would like to emphasize that no reduction has been made to derive
the Schr\"odinger-like equation, so that all relativistic corrections
are included in our calculation.
The calculated  meson-nucleus bound state energies for $^{208}$Pb,
are listed in Table~\ref{energy}.


The results show that both the $D^-$ and $\bar{D}^0$ are bound in $^{208}$Pb
with the usual $\omega$ coupling constant.
For the $D^-$ the Coulomb force
provides roughly 24 MeV of  binding for the $1s$ state, and is strong enough
to bind the system even with the much more repulsive $\omega$ coupling
(viz., $1.4^2V_\omega^q$).
The  $\d0bar$ with the stronger $\omega$ coupling is not bound.
Note that the difference between $\d0bar$ and $D^-$ without the Coulomb force
 is due to the interaction with the
$\rho$ meson, which is attractive for the $\d0bar$ but repulsive for
the $D^-$. For completeness, we also calculated the binding energies
for the $D^0$, which is deeply bound since the $\omega$ interaction
with the light antiquarks is attractive. However, the expected
large width associated with strong absorption may render it
experimentally inaccessible. It is an extremely important experimental
challenge to see whether it can be detected.


The eigenfunctions for the Schr\"odinger-like equation are shown
in  Fig.~\ref{dmeswf}, together with the baryon density distribution
in $^{208}$Pb.
For the usual $\omega$ coupling, the eigenstates ($1s$ and $1p$)
are well within the nucleus, and behave as expected at the origin.
For the stronger $\omega$ coupling, however, the $D^-$ meson
is considerably pushed out of the nucleus. In this case,
the bound state (an atomic state) is formed
solely due to the Coulomb force.
An experimental determination of whether this is a nuclear state or an atomic
state would give a strong constraint on the $\omega$ coupling.
We note, however, that
because it is very difficult to produce $D$-mesic nuclei with
small momentum transfer, and the $D$-meson production
cross section is small compared with the background from other channels,
it will be a challenging task to detect such bound states
experimentally~\cite{hayaex}.

To summarize, we have calculated the $_{D^0}^{208}$Pb,
$_{\d0bar}^{208}$Pb and $_{D^-}^{208}$Pb
bound state energies in QMC.
The potentials for the mesons were calculated
self-consistently in local density approximation.
In spite of possible model-dependent uncertainties, our results suggest that 
the $D^-$ meson should be bound in $^{208}$Pb due to two quite
different mechanisms, namely, the scalar and attractive
$\sigma$ mean field even without the assistance of the Coulomb force
in the case of the normal vector potential ($V^q_\omega$), and 
solely due to the Coulomb force in the case of the stronger 
vector potential ($\tilde{V}^q_\omega$). 
(We recall that the kaon is a pseudo-Goldstone boson 
and expected to be difficult to treat properly with the usual bag model. 
Thus, the analysis of Ref.~\cite{kaon} on the vector potential for the light 
quarks inside the kaon bag may not be applicable to the light 
quarks inside the D-meson.)  

Thus, whether or not the $\d0bar$-$^{208}$Pb bound states exist
would give new information as to whether the interactions of
light quarks in a heavy meson are the same as those in a nucleon.
The enormous difference
between the binding energies of the $D^0$ ($\sim
100$ MeV) and the $\d0bar$ ($\sim 10$ MeV) is a simple consequence of the
presence of a strong Lorentz vector mean-field, while the existence of
any binding at all would give us important information concerning the
role of the Lorentz scalar $\sigma$ field (and hence dynamical symmetry
breaking) in heavy quark systems. In spite of the perceived experimental
difficulties,
we feel that the search for these bound systems should have a very high
priority.

\noindent{\bf Acknowledgment}\\
We would like to thank R.S. Hayano for useful discussions concerning
the experimental possibilities for detecting $D$-mesic bound states.
This work was supported by the Australian Research Council.

%

\newpage
%
\begin{table}[htb]
\begin{center}
\caption{
The physical masses fitted in free space,
the bag parameters, $z$, and the bag radii in free space, $R$.
The quantities with an asterisk, $^*$, are those quantities calculated
at normal nuclear matter density, $\rho_0 = 0.15$ fm$^{-3}$.
They are obtained with the bag constant, $B = (170.0$ MeV$)^4$, current
quark masses, $m_u = m_d = 5$ MeV  and
$m_c = 1300$ MeV. (Note that the free
space widths for $D$ and $\bar{D}$ mesons are negligible~\protect\cite{pdata}.)
}
\label{bagparam}
\begin{tabular}[t]{llclll}
&mass (MeV) &$z$ &$R$ (fm)& $m^*$ (MeV) & $R^*$ (fm)\\
\hline
$N$            &939.0 (input) &3.295 &0.800 (input) &754.5  &0.786\\
$D,\bar{D}$   &1866.9 (input)&1.389 &0.731         &1804.9 &0.730\\
%
\end{tabular}
\end{center}
\end{table}
%
\begin{table}[htb]
\begin{center}
\caption{
Calculated $D^-$, $\d0bar$ and $D^0$ meson bound state energies (in MeV)
in $^{208}$Pb for different potentials.
The widths for the mesons are all set to zero, both
in free space and inside $^{208}$Pb. Note that the $D^0$ bound states
energies calculated with $\tilde{V}^q_\omega$ will be much larger than
those calculated with $V^q_\omega$ (in absolute value).
}
\label{energy}
\begin{tabular}[t]{lccc|ccc}
state  &$D^- (\tilde{V}^q_\omega)$ &$D^- (V^q_\omega)$
&$D^- (V^q_\omega$, no Coulomb) &$\d0bar (\tilde{V}^q_\omega)$
&$\d0bar (V^q_\omega)$ &$D^0 (V^q_\omega)$ \\
\hline
                         1s &-10.6 &-35.2 &-11.2 &unbound &-25.4 &-96.2\\
                         1p &-10.2 &-32.1 &-10.0 &unbound &-23.1 &-93.0\\
                         2s & -7.7 &-30.0 & -6.6 &unbound &-19.7 &-88.5\\
\end{tabular}
\end{center}
\end{table}
\newpage
%
\begin{figure}[hbt]
\begin{center}
\epsfig{file=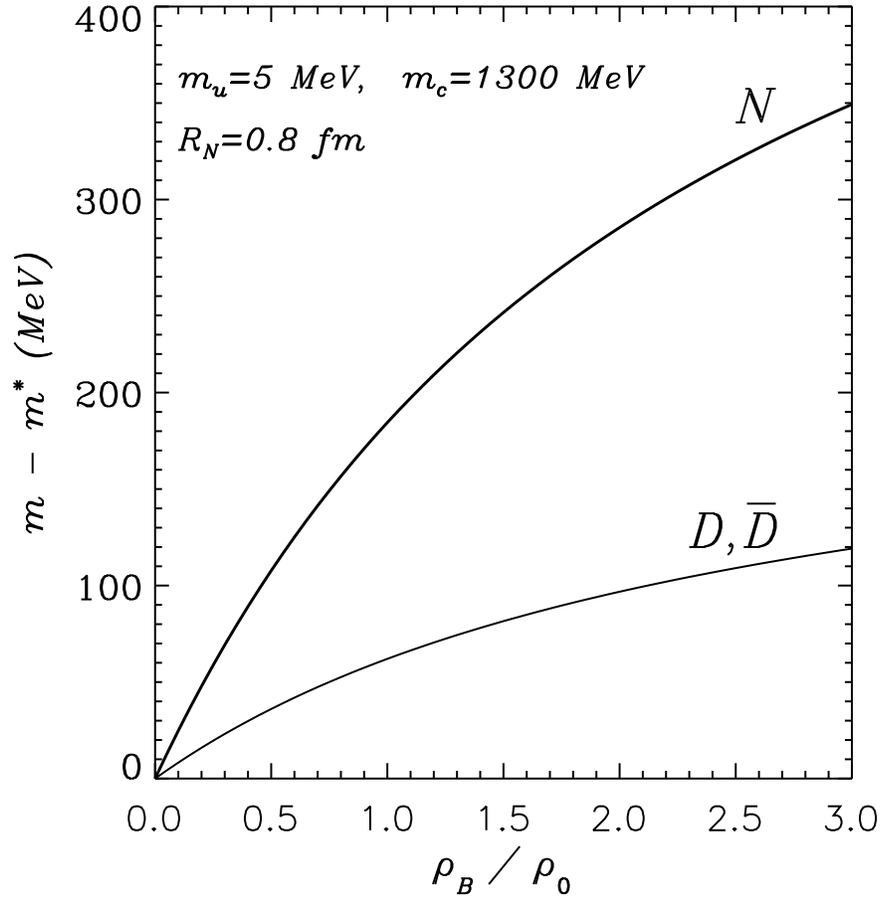,height=13cm}
\caption{Shift of the mass of the nucleon and the
$D$ or \protect{$\bar{D}$} meson.
(Normal nuclear matter density, $\rho_0$, is 0.15 fm$^{-3}$.)}
\label{dmassmatter}
\end{center}
\end{figure}
\newpage
%
\begin{figure}[hbt]
\begin{center}
\epsfig{file=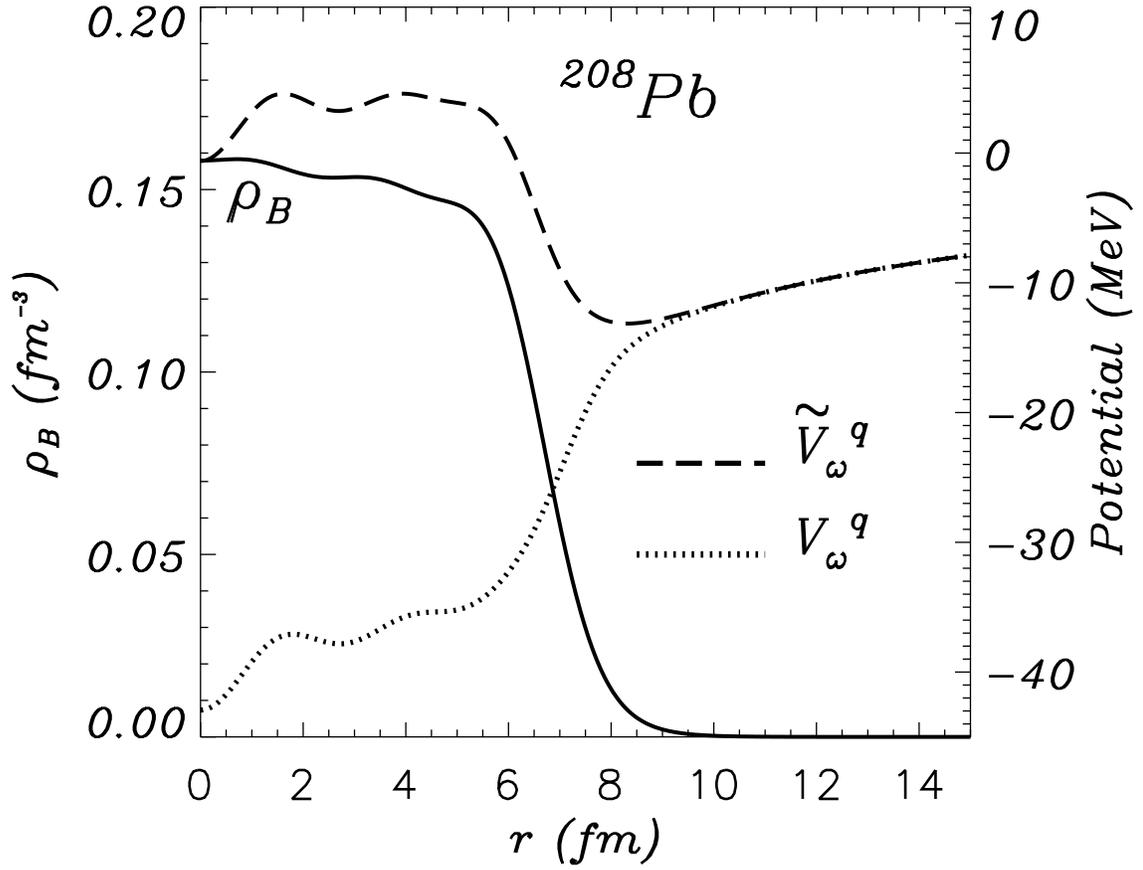,height=13cm}
\caption{Sum of the scalar, vector and Coulomb potentials for the $D^-$
meson in $^{208}$Pb for two cases,
$(m^*_{D^-}(r) - m_{D^-}) + \tilde{V}^q_\omega(r)
+ \frac{1}{2} V^q_\rho(r) - A(r)$ (the dashed line) and
$(m^*_{D^-}(r) - m_{D^-}) + V^q_\omega(r)
+ \frac{1}{2} V^q_\rho(r) - A(r)$ (the dotted line), 
where $\tilde{V}^q_\omega(r) = 1.4^2 V^q_\omega(r)$.
}
\label{dmespot}
\end{center}
\end{figure}
\newpage
%
\begin{figure}[hbt]
\begin{center}
\epsfig{file=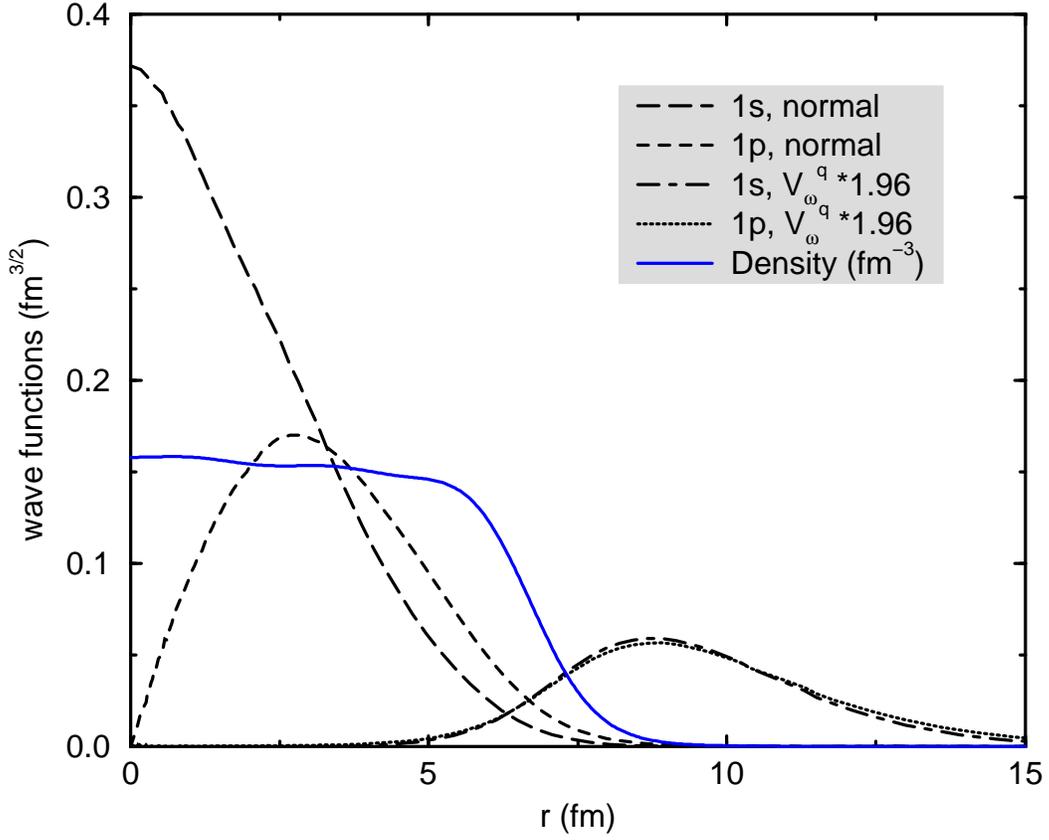,height=13cm}
\caption{The Schr\"odinger-like bound state wave functions of the $D^-$ meson
in $^{208}$Pb, for two different $\omega$ meson coupling strengths.
See also the caption of Fig.~\protect\ref{dmespot}.
The wavefunction is normalized as follows:
$\int_0^{\infty}\!dr\,4\pi r^2 |\Phi(r)|^2 = 1$.
}
\label{dmeswf}
\end{center}
\end{figure}
%
%
\end{document}